\documentclass[letterpaper,dvipdfm,epsf,11pt]{iopart}
\eqnobysec
\usepackage{iopams,amsfonts,rotating,graphicx,fancybox,fancyhdr}
\usepackage{eucal,dsfont}
\newcommand{\bm}{\boldsymbol}

\newcommand{\mkbox}[3]{\hbox{\vrule
      \vbox to  #1{\hrule \vss
                  \hbox to #2{\hss#3\hss}\vss
                  \hrule}\vrule}}

\begin{document}

\title[Replica Approach in Random Matrix Theory]
        {Replica Approach in Random Matrix Theory \footnote[8]{\noindent {\footnotesize Chapter for \textsc{The
Oxford Handbook of Random Matrix Theory}, edited by G.~Akemann,
J.~Baik, and P.~Di Francesco, to be published by Oxford University
Press.}}}

\author{Eugene Kanzieper}

\address{Department of Applied Mathematics, H.I.T. -- Holon Institute of
    Technology\\ Holon 58102,
    Israel} \eads{
    \mailto{Eugene.Kanzieper@hit.ac.il}}

\begin{abstract}
This Chapter outlines the replica approach in Random Matrix Theory.
Both fermionic and bosonic versions of the replica limit are
introduced and its trickery is discussed. A brief overview of early
heuristic treatments of zero-dimensional replica field theories is
given to advocate an exact approach to replicas. The latter is
presented in two elaborations: by viewing the $\beta=2$ replica
partition function as the Toda Lattice and by embedding the replica
partition function into a more general theory\linebreak of $\tau$
functions.\\ \\
\texttt{arXiv:~0909.3198v2}
\end{abstract}

\newpage
\tableofcontents
\newpage
\section{Introduction} \label{Sec1}

\subsection{Resolvent as a field integral}
In physics of disorder, all observables depend in highly nonlinear
fashion on a stochastic Hamiltonian hereby making a nonperturbative
calculation of their ensemble averages very difficult. To determine
the average quantities in an interactionless system, one has to know
the spectral statistical properties of a single particle Hamiltonian
${\boldsymbol {\mathcal H}}$ contained in the mean product of
resolvents $G(z) = {\rm tr} \left( z - {\boldsymbol {\mathcal H}}
\right)^{-1}$ defined for a generic, complex valued argument $z \in
{\mathbb C} \setminus {\mathbb R}$. Each of the resolvents can
exactly be represented as a ratio of two functional integrals
running over an auxiliary vector field ${\boldsymbol \psi}$ which
may consist of {\it either} commuting (bosonic, ${\boldsymbol
\psi}={\boldsymbol s}$) {\it or} anticommuting (fermionic,
${\boldsymbol \psi}={\boldsymbol \chi}$) entries.

In the random matrix theory (RMT) limit, when a system Hamiltonian
(or a scattering matrix) is modelled by an $N \times N$ random
matrix ${\boldsymbol {\mathcal H}}$ of prescribed symmetry, the
resolvent $G(z)$ equals
\begin{eqnarray} \fl
    \label{ratio}
    G(z)
    =
    i \eta\, {\mathfrak s}_z \int {\cal D}[{\boldsymbol {\bar \psi}},{\boldsymbol \psi}] \;
    {\bar \psi}_\ell \psi_\ell \, e^{
    i {\mathfrak s}_z\, {\mathcal S}_{{\boldsymbol {\mathcal H}}}(z;\, {\boldsymbol {\bar \psi}}, {\boldsymbol \psi})}
    \left(
    \int {\cal D}[{\boldsymbol {\bar \psi}},{\boldsymbol \psi}] \;
    e^{
    i {\mathfrak s}_z\, {\mathcal S}_{{\boldsymbol {\mathcal H}}}(z; \, {\boldsymbol {\bar \psi}}, {\boldsymbol \psi})
    }
    \right)^{-1}.
\end{eqnarray}
Here, $
    {\mathcal S}_{{\boldsymbol {\mathcal H}}}=
    {\boldsymbol {\bar \psi}} (z-{\boldsymbol{\mathcal H}}){\boldsymbol {\psi}}
    ={\bar \psi}_j
    \left( z\, \delta_{jk}-{\cal H}_{jk} \right)
    \psi_{k}$ (summation over repeated Latin indices is
    assumed throughout this Section); vector ${\boldsymbol \psi}$ is defined by ${\boldsymbol \psi} = (\psi_1,\cdots,
    \psi_N)^{\rm T}$, whilst ${\boldsymbol {\bar \psi}}=(\bar{\psi}_1,\cdots, {\bar \psi}_N)$ is its proper conjugate. The parameter $\eta$
accounts for the nature of both vectors: it is set
    to $+1$ for fermions (${\boldsymbol \psi}={\boldsymbol \chi}$) and to $-1$ for bosons (${\boldsymbol
    \psi}={\boldsymbol s}$). In the latter case, convergence of
    field integrals is ensured by the regularizer ${\mathfrak s}_z ={\rm sgn}(\mathfrak{Im\,}z)$. The notation ${\cal D}[{\boldsymbol {\bar \psi}},{\boldsymbol
    \psi}]$ stands for the integration measure~\footnote[1]{Integrals over anticommuting (Grassmann) variables
    $\chi$ and ${\bar \chi}$ are normalised according to
\begin{eqnarray}
\int d\chi\,
\chi = \int d\bar{\chi}\, \bar{\chi}= (2\pi)^{1/2}. \nonumber
\end{eqnarray}} ${\cal D}[{\boldsymbol {\bar \psi}},{\boldsymbol
    \psi}] = (2\pi)^{-N}\prod_{j=1}^N d{\bar \psi}_j d\psi_j$. 
    
Equation (\ref{ratio}) may conveniently be viewed as a
    consequence of the identity
\begin{eqnarray}
\label{G-f-log}
    G(z) = \eta \frac{\partial}{\partial z} \log\, \det{}^{\eta} (z-{{\boldsymbol {\mathcal H}}}),
\end{eqnarray}
combined with the field integral representations of the determinant
($\eta=+1$, ${\boldsymbol \psi}={\boldsymbol \chi}$) and/or its
inverse ($\eta=-1$, ${\boldsymbol \psi}={\boldsymbol s}$):
\begin{eqnarray} \label{det}
    \det{}^\eta (z-{{\boldsymbol {\mathcal H}}}) = (i\eta \,\mathfrak{s}_z)^N
    \int {\cal D}[{\boldsymbol {\bar \psi}},{\boldsymbol \psi}] \;
     e^{
    i {\mathfrak s}_z\, {\mathcal S}_{{\boldsymbol {\mathcal H}}}(z;\, {\boldsymbol {\bar \psi}}, {\boldsymbol \psi})}.
\end{eqnarray}
Although exact, both bosonic and fermionic versions of
Eq.~(\ref{ratio}) are a bit too inconvenient for a nonperturbative
ensemble averaging due to the awkward random denominator.

To get rid of it, several field theoretic frameworks have been
devised by theoretical physicists: the replica trick
\cite{W-1979,SW-1980,ELK-1980}, the supersymmetry method
\cite{E-1982a,E-1982b}, and the dynamic (Keldysh) approach
\cite{HS-1990,KA-1999}. Leaving aside the supersymmetry and the
Keldysh techniques \footnote[2]{An introductory exposition of the
supersymmetry method can be found in Chapter 7 of this Handbook as well as in the earlier review papers
\cite{E-1983,VWZ-1985} and the monograph \cite{E-1997}. For a review
of the Keldysh approach, the reader is referred to \cite{KL-2009};
see also the paper \cite{AK-2000} where the Keldysh technique is
discussed in the RMT context.}, this Chapter aims to provide an
elementary introduction to notoriously known replica approach whose
legitimacy has been a point of controversy
\cite{VZ-1985,KM-1999b,Z-1999} for over two decades. Recently
discovered integrability \cite{K-2002,K-2005,OK-2007} of
zero-dimensional replica field theories will be a dominant motif of
this contribution.

\subsection{How replicas arise and why they are tricky}
Both fermionic and bosonic replicas are based on the
identity~\footnote[3]{Mostly known in condensed matter physics
community since the paper by Edwards and Anderson \cite{EA-1975} on
spin glasses (see also \cite{E-1975}), the recipe of calculating the
average of a logarithm based on Eq.~(\ref{log-id}) dates at least as
far back as 1934, see the book \cite{HLP-1934} by Hardy, Littlewood
and P\'olya. }
\begin{eqnarray}
\label{log-id}
    \log {\mathfrak X} = \lim_{n\rightarrow 0} \frac{{\mathfrak X}^n-1}{n}
\end{eqnarray}
which can be very useful in evaluating the average of a logarithm
$\left< \log {\mathfrak X}\right>$ of the random variable
${\mathfrak X}$. Indeed, identifying ${\mathfrak X}$ with
$\det{}^\eta (z-{{\boldsymbol {\mathcal H}}})$ in
Eq.~(\ref{G-f-log}), and further combining Eq.~(\ref{log-id}) with
the field integral representation Eq.~(\ref{det}) of the (inverse)
determinant, we {\it formally} relate the resolvent
\begin{eqnarray}
\label{G-replicas-n}
    G(z) = \eta \lim_{n\rightarrow 0} \frac{1}{n} \frac{\partial}{\partial z}
    \,{\mathcal Z}_n^{(\eta)}(z)
\end{eqnarray}
to the `partition function'
\begin{eqnarray} \label{G-replicas-n-2}
    {\mathcal Z}_n^{(\eta)}(z) = \prod_{\alpha=1}^n
    \int {\cal D}[{\boldsymbol {\bar \psi}}^{(\alpha)},{\boldsymbol \psi}^{(\alpha)}] \;
    \exp\left[i\mathfrak{s}_z\,
    {\boldsymbol {\bar \psi}}^{(\alpha)} (z-{\boldsymbol{\mathcal H}}){\boldsymbol {\psi}}^{(\alpha)}\right]
\end{eqnarray}
of $n$ copies, or replicas, of the initial random system. The pair
of formulae Eqs.~(\ref{G-replicas-n}) and (\ref{G-replicas-n-2}),
known as the {\it replica trick}~\footnote[4]{The alert reader might
already detect some trickery behind Eqs.~(\ref{G-replicas-n}) and
(\ref{G-replicas-n-2}). Their mathematical status will be clarified
in Section \ref{trickery-sec}.}, achieve our goal of removing a
random denominator from Eq.~(\ref{ratio}) and hint that a
nonperturbative calculation of the mean product of resolvents may
become feasible.

In order to keep the discussion concrete and set up notation, we
further assume that the matrix Hamiltonian ${\boldsymbol {\mathcal
H}}$ is drawn from the paradigmatic Gaussian Unitary Ensemble (GUE)
associated with the probability measure~\footnote{See also Chapter 4 of this Handbook which outlines the method of orthogonal polynomials for unitary invariant random matrix models.} \cite{M-2004}
\begin{eqnarray}
\label{gue-prob}
    P_N[{\boldsymbol {\mathcal H}}] \,{\mathcal D}{\boldsymbol {\mathcal H}} =
    \pi^{-N^2/2} \exp(-{\rm tr\,} {\boldsymbol {\mathcal H}}^2)
    \prod_{j=1}^N d{\mathcal H}_{jj} \prod_{1\le j<k}^N d{\mathcal H}_{jk} d\bar{{\mathcal H}}_{jk}.
\end{eqnarray}
Then, the resolvent $g(z) = \left< G(z) \right>$ averaged with
respect to the GUE probability measure should be furnished by the
limiting procedure
\begin{eqnarray}
\label{av-res}
    g(z) =\eta \lim_{n\rightarrow 0} \frac{1}{n} \frac{\partial}{\partial z}
    \,  \langle {\mathcal Z}_n^{(\eta)}
(z)\rangle,
\end{eqnarray}
where the (average) {\it replica partition function} equals
\begin{eqnarray}
\label{rpf-av}
    \langle {\mathcal Z}_n^{(\eta)} (z)\rangle = \prod_{\alpha=1}^n
    \int {\cal D}[{\boldsymbol {\bar \psi}}^{(\alpha)},{\boldsymbol \psi}^{(\alpha)}] \;
    \exp\left(-i\eta\,\mathfrak{s}_z\,z\,{\rm tr\,} {\boldsymbol \sigma}
    +\frac{\eta}{4} {\rm tr\,}{\boldsymbol \sigma}^2\right).
\end{eqnarray}
Here, the Hermitean matrix ${\boldsymbol \sigma}={\boldsymbol
\sigma}^\dagger$ acting in the replica space is defined as
\begin{eqnarray}
    \sigma_{\alpha\beta} = \left( \psi_j \otimes {\bar \psi}_j\right)_{\alpha\beta}=
    {\psi}_j^{(\alpha)}{\bar{\psi}}_j^{(\beta)},\;\;\; \alpha,\beta \in (1,\dots,n).
\end{eqnarray}
Notice that ensemble averaging of ${\mathcal Z}_n^{(\eta)}(z)$
[Eq.~(\ref{G-replicas-n-2})] has induced an effective ${\boldsymbol
{\psi}}^4$ interaction between $n$ replicated random systems as
described by the interaction term ${\rm tr\,}{\boldsymbol
\sigma}^2={\rm tr\,}[(\psi_j \otimes {\bar \psi}_j)^2]$ in the
action. To make further progress in evaluating $\langle {\mathcal
Z}_n^{(\eta)} (z)\rangle$, the interaction term may routinely be
decoupled by means of the Hubbard-Stratonovich transformation
\cite{E-1997,VWZ-1985}.

\subsubsection{Fermionic replicas}\noindent\newline\newline
In case of fermionic fields (${\boldsymbol \psi} = {\boldsymbol
\chi}$, $\eta=+1$), the ${\boldsymbol \chi}^4$ interaction can be
decoupled via $n\times n$ Hermitean matrix field ${\boldsymbol
{\mathcal Q}}_n$ as
\begin{eqnarray}
    \exp\left(
        +\frac{1}{4} {\rm tr\,} {\boldsymbol {\sigma}}^2
    \right) = \pi^{-n^2/2}
    \int_{{\boldsymbol {\mathcal Q}}_n^\dagger = {\boldsymbol {\mathcal Q}}_n}
    {\mathcal D}{\boldsymbol {\mathcal Q}}_n
    \exp\left[-{\rm tr} \left( {\boldsymbol {\mathcal Q}}_n^2
    + {\boldsymbol \sigma} {\boldsymbol {\mathcal Q}}_n \right) \right].
\end{eqnarray}
Performing the integration over fermionic fields in
Eq.~(\ref{rpf-av}) with the help of Eq.~(\ref{det}),
\begin{eqnarray}
\label{chi-int} \fl
    \prod_{\alpha=1}^n
    \int {\cal D}[{\boldsymbol {\bar \chi}}^{(\alpha)},{\boldsymbol \chi}^{(\alpha)}] \;
    \exp\left(-i\mathfrak{s}_z\,z\,{\rm tr\,} {\boldsymbol \sigma}
    +\frac{1}{4} {\rm tr\,}{\boldsymbol \sigma}^2\right) \nonumber \\
    =
    \pi^{-n^2/2}(-i{\mathfrak s}_z)^{nN}
    \int_{{\boldsymbol {\mathcal Q}}_n^\dagger = {\boldsymbol {\mathcal Q}}_n}
    {\mathcal D}{\boldsymbol {\mathcal Q}}_n
    \,e^{-{\rm tr\,} {\boldsymbol {\mathcal Q}}_n^2}\,
    {\rm det}{}^N (z - i{\mathfrak s}_z {\boldsymbol {\mathcal Q}}_n),
\end{eqnarray}
we arrive at the {\it fermionic} replica limit
\begin{eqnarray}
    \label{fermi-res-av}
    g(z)
    = \lim_{n \rightarrow 0}\, \frac{1}{n}\frac{\partial}{\partial z}
    \,\langle{\mathcal Z}_n^{\rm{(+)}}(z)\rangle
\end{eqnarray}
that relates the average resolvent $g(z)$ to the {\it fermionic}
replica partition function
\begin{eqnarray}
\label{fer-pf}
    \langle {\mathcal Z}_n^{\rm{(+)}}(z)\rangle = \int_{{\boldsymbol {\mathcal Q}}_n^\dagger= {\boldsymbol {\mathcal Q}}_n}
    {\cal D}{\boldsymbol {\mathcal Q}}_n\, \,e^{-{\rm tr\,} {\boldsymbol {\mathcal Q}}_n^2}\,
    {\rm det}{}^{N} \left(
        z - i{\mathfrak s}_z  {\boldsymbol {\mathcal Q}}_n
    \right).
   \end{eqnarray}
Before discussing this result, let us turn to the bosonic version of
the replica trick.

\subsubsection{Bosonic replicas}\noindent\newline\newline
In case of bosonic fields (${\boldsymbol \psi} = {\boldsymbol s}$,
$\eta=-1$), decoupling of the ${\boldsymbol s}^4$ interaction can be
carried out in a similar fashion. Making use of yet another variant
of the Hubbard-Stratonovich transformation
\begin{eqnarray}
    \exp\left(
        -\frac{1}{4} {\rm tr\,} {\boldsymbol {\sigma}}^2
    \right) = \pi^{-n^2/2}
    \int_{{\boldsymbol {\mathcal Q}}_n^\dagger = {\boldsymbol {\mathcal Q}}_n}
    {\mathcal D}{\boldsymbol {\mathcal Q}}_n
    \exp\left[-{\rm tr} \left( {\boldsymbol {\mathcal Q}}_n^2
    + i {\boldsymbol \sigma} {\boldsymbol {\mathcal Q}}_n \right)\right],
\end{eqnarray}
and integrating out bosonic fields in Eq.~(\ref{rpf-av}) with the
help of Eq.~(\ref{det}),
\begin{eqnarray}
\label{s-int} \fl
    \prod_{\alpha=1}^n
    \int {\cal D}[{\boldsymbol {\bar s}}^{(\alpha)},{\boldsymbol s}^{(\alpha)}] \;
    \exp\left(+i\mathfrak{s}_z\,z\,{\rm tr\,} {\boldsymbol \sigma}
    -\frac{1}{4} {\rm tr\,}{\boldsymbol \sigma}^2\right) \nonumber \\
    =
    \pi^{-n^2/2}(+i{\mathfrak s}_z)^{nN}
    \int_{{\boldsymbol {\mathcal Q}}_n^\dagger = {\boldsymbol {\mathcal Q}}_n}
    {\mathcal D}{\boldsymbol {\mathcal Q}}_n
    \,e^{-{\rm tr\,} {\boldsymbol {\mathcal Q}}_n^2}\,
    {\rm det}{}^{-N} (z - {\mathfrak s}_z {\boldsymbol {\mathcal Q}}_n),
\end{eqnarray}
we express the average resolvent
\begin{eqnarray}
    \label{bose-res-av}
    g(z)
    = - \lim_{n \rightarrow 0}\, \frac{1}{n}\frac{\partial}{\partial z}
    \,\langle{\mathcal Z}_n^{\rm{(-)}}(z)\rangle
\end{eqnarray}
through the {\it bosonic} replica partition function
\begin{eqnarray}
\label{bos-pf}
    \langle {\mathcal Z}_n^{\rm{(-)}}(z)\rangle =
    \int_{{\boldsymbol {\mathcal Q}}_n^\dagger= {\boldsymbol {\mathcal Q}}_n}
    {\cal D}{\boldsymbol {\mathcal Q}}_n\, \,e^{-{\rm tr\,} {\boldsymbol {\mathcal Q}}_n^2}\,
    {\rm det}{}^{-N} \left(
        z - {\mathfrak s}_z  {\boldsymbol {\mathcal Q}}_n
    \right).
   \end{eqnarray}

\subsubsection{Subtleties of the replica limit}\label{trickery-sec}\noindent\newline\newline
Seemingly innocent at first glance, both fermionic
[Eqs.~(\ref{fermi-res-av}) and (\ref{fer-pf})] and bosonic
[Eqs.~(\ref{bose-res-av}) and (\ref{bos-pf})] replica prescriptions
appear to be counterintuitive and rising fundamental mathematical
questions \cite{P-2003}. Indeed, due to a particular integration
measure which makes no sense for $n$ other than positive integers
($n\in {\mathbb Z}^+$), the matrix-integral representation of
average replica partition functions
\begin{eqnarray}
\label{rpf}
    \langle {\mathcal Z}_n^{\rm{(\eta)}}(z)\rangle =
    \int_{{\boldsymbol {\mathcal Q}}_n^\dagger= {\boldsymbol {\mathcal Q}}_n}
    {\cal D}{\boldsymbol {\mathcal Q}}_n\, \,e^{-{\rm tr\,} {\boldsymbol {\mathcal Q}}_n^2}\,
    {\rm det}{}^{\,\eta N} \big(
        z - \sqrt{i}^{\,1+\eta}{\mathfrak s}_z  {\boldsymbol {\mathcal Q}}_n
    \big)
\end{eqnarray}
cannot directly be used to implement the replica limit
Eqs.~(\ref{av-res}) as the latter is determined by the behaviour of
$\langle {\mathcal Z}_n^{\rm{(\eta)}}(z)\rangle$ in a close {\it
vicinity} of $n=0$. This mismatch between the `available' ($n\in
{\mathbb Z}^+$) and the `needed' ($n\in {\mathbb R}$) is at the
heart of the trickery the replica field theories have often been
charged \cite{VZ-1985,Z-1999}.

The canonical way to bridge this gap is to determine the average
replica partition functions $\langle {\mathcal
Z}_n^{\rm{(\eta)}}(z)\rangle$ for $n \in {\mathbb Z}^+$, and then
attempt to analytically continue them to $n\in {\mathbb R}$, in
general, and to a vicinity of $n=0$, in particular. This is a
nontrivial task for two major reasons: (i) The analytic continuation
of the replica partition function away from $n$ integers should not
necessarily be unique~\footnote[5]{For entire functions, the
uniqueness is guaranteed by a boundedness property as formulated by
Carlson's theorem \cite{T-1932}.}. (ii) To retain control over the
analytic continuation, the latter must rest on an exact calculation
of the average replica partition function for $n \in {\mathbb Z}^+$.
Early approaches to replica field theories seem to underestimate
these two points bringing a number of pathological results even in
the RMT setting (see Section \ref{early-st}).

\section{Early studies: Heuristic approach to replicas}
\label{early-st}

Exact evaluation of replica partition functions, whilst welcomed, is
quite a challenge. At the same time, their approximate calculation
is often feasible in a certain region of parameter space where a
saddle point procedure can be justified. In doing so, one is
naturally led to the {\it `replica symmetric'} and {\it `replica
asymmetric'} saddle point manifolds as discussed below.

\subsection{Density of eigenvalues in the GUE}
In the RMT context, a saddle point evaluation of the replica
partition function Eq.~(\ref{rpf}) makes sense if the dimension $N$
of the random matrix ${{\boldsymbol {\mathcal H}}}$ is large enough.
For not too large replica parameter $n\in{\mathbb Z}^+$ (in
particular, $n$ should not scale with $N$), the dominating
contribution to $\langle {\mathcal Z}_n^{\rm{(\eta)}}(z)\rangle$ is
expected to come from the configurations ${\boldsymbol {\mathcal
Q}}_n^{{\rm (sp)}}$ determined by the saddle point equation
\begin{eqnarray}
    \label{sp-eq}
    \frac{\delta}{\delta {\boldsymbol {\mathcal Q}}_n} \, {\rm tr}
        \Big[
        {\boldsymbol {\mathcal Q}}_n^2 - \eta N \, \log (z - \sqrt{i}^{\,1+\eta}
        {\mathfrak s}_z {\boldsymbol {\mathcal Q}}_n)
    \Big]
    =0.
\end{eqnarray}
Its solutions form $2^{n}$ saddle point manifolds
\begin{eqnarray}
    \label{sp-solution}
    {\boldsymbol {\mathcal Q}}_n^{\rm (sp)} = \frac{{\mathfrak s}_z}{\sqrt{i}^{\, 1+\eta}} \sqrt{\frac{N}{2}}
    \,{\rm diag}(e^{i \kappa_1\theta},\dots, e^{i \kappa_{n} \theta})
\end{eqnarray}
with $\kappa_\ell$ taking on the values $\pm 1$ independently of
each other. Here,
\begin{eqnarray}
e^{i\theta}=z_{\rm s} + i \sqrt{1-z_{\rm s}^2},
\end{eqnarray}
where $z_{\rm s}$ stands for the scaled energy $z_{\rm
s}=z/{\mathcal D}_{\rm edge}$ with ${\mathcal D}_{\rm
edge}=\sqrt{2N}$ being the endpoint of the spectrum support. (Hence,
the spectrum bulk is situated within the segment $|\mathfrak{Re\,}
z_{\rm s}|<1$.)

\subsubsection{Bosonic replicas}\noindent\newline\newline
Out of the plethora of saddles Eq.~(\ref{sp-solution}), only the
distinguished {\it replica symmetric} manifold
\begin{eqnarray}
    \label{sp-boson-symmetric}
    {\boldsymbol {\mathcal Q}}_n^{\rm (sp)} \Big|_{{\rm sym}}
    =  {\mathfrak s}_z\, \sqrt{\frac{N}{2}}
     \,e^{i \theta} \,\otimes
        \mathds{1}_{n}
\end{eqnarray}
contributes the bosonic replica partition function
Eq.~(\ref{bos-pf}). This is so because
Eq.~(\ref{sp-boson-symmetric}) is the only saddle \cite{Z-1999} (i)
reachable by continuous deformation of the integration contour in
Eq.~(\ref{bos-pf}) without crossing the hypersurface defined by the
singularities of ${\rm det}{}^{-N}(z-{\mathfrak s}_z {\boldsymbol
{\mathcal Q}}_n)$ and (ii) compatible with analyticity of
the average resolvent Eq.~(\ref{bose-res-av}) at infinity. In the
leading order in the large parameter $N$, the bosonic replica
partition function is then approximated by
\begin{eqnarray} \fl
    \label{BPF-sp}
    \langle {\mathcal Z}_n^{\rm{(-)}}(z=\epsilon -i0)\rangle \simeq \left(  \frac{N}{2} \right)^{nN/2}
    \, (2 \sin \, \theta)^{-n^2/2}  \exp\left[ \frac{nN}{2} \left( e^{2i\theta} -2i\theta  \right)
     - i \, \frac{n^2}{2}\,\left(\theta - \frac{\pi}{2}\right) \right]. \nonumber\\
     {}
\end{eqnarray}

By derivation, Eq.~(\ref{BPF-sp}) holds for $n \in {\mathbb Z}^+$.
To retrieve the average density of eigenlevels
\begin{eqnarray} \label{doe}
\varrho(\epsilon)=\frac{1}{\pi} \mathfrak{Im\,} g(\epsilon-i0)
\end{eqnarray}
through the replica limit Eq.~(\ref{bose-res-av}), one should
analytically continue $\langle {\mathcal Z}_n^{\rm{(-)}}(z)\rangle$
away from $n \in {\mathbb Z}^+$. To be on the safe side, such an
analytic continuation must rest on an exact integer-$n$ result for
$\langle {\mathcal Z}_n^{\rm{(-)}}(z)\rangle$. The latter is sadly
unavailable. Not being spoilt for choice, one could try to
analytically continue the bosonic replica partition function to the
domain $n \in {\mathbb R}^+$ taking the approximate result
Eq.~(\ref{BPF-sp}) as a starting point and merely assuming it to
hold, as it stands, in the right vicinity of $n =0$. Then, the
replica limit Eq.~(\ref{bose-res-av}) can be taken to yield the
Wigner semicircle \cite{EW-1980}
\begin{eqnarray}
    \label{WSC-b=2}
    \varrho(\epsilon_{\rm s}) = \frac{2}{\pi}
      \sqrt{1 - \epsilon_{\rm s}^2}, \;\;\; |\epsilon_{\rm s}|\le 1.
\end{eqnarray}
Here, $\epsilon_{\rm s} = \epsilon/{\mathcal D}_{\rm edge}$. The
$1/N$ correction to Eq.~(\ref{WSC-b=2}) can be obtained in a similar
fashion and is known to vanish within the replica symmetric ansatz.

Similarly, replica symmetric saddle point calculations performed for
the Gaussian Orthogonal Ensemble (GOE) and the Gaussian Symplectic
Ensemble (GSE) of random matrices yield
\cite{EJ-1976,VZ-1984,DJ-1990,IMS-1997}
\begin{eqnarray} \fl
    \label{goegse-wigner}
        \varrho(\epsilon_{\rm s}) = \frac{2}{\pi}
      \sqrt{1 - \epsilon_{\rm s}^2} + \frac{1}{2\pi N} \left(
        1-\frac{2}{\beta}
      \right) \left[
        \frac{1}{\sqrt{1-\epsilon_{\rm s}^2}} - \pi \delta(\epsilon_{\rm s}^2-1)
      \right], \;\;\; |\epsilon_{\rm s}|\le 1.
\end{eqnarray}
Here, $\beta$ is the Dyson symmetry index \cite{M-2004} taking the
values $\beta=1$ for GOE, $\beta=2$ for GUE, and $\beta=4$ for GSE.

Importantly, both leading and subleading in $1/N$ terms in
Eq.~(\ref{goegse-wigner}) for the average densities of eigenlevels
contain no terms oscillating on the scale of the mean level spacing.
Since the replica asymmetric saddles are inaccessible as explained
on general grounds \cite{Z-1999} in (i) and (ii) below
Eq.~(\ref{sp-boson-symmetric}), we are led to conclude that the
saddle point approach to bosonic replicas fails to reproduce truly
non-perturbative features of the eigenlevel density.

\subsubsection{Fermionic replicas}\noindent\newline\newline
Looking for some insight into a possible r\^ole played by {\it
replica asymmetric} saddles, we turn to the approximate performance
of fermionic replicas Eq.~(\ref{fer-pf}). In this case, all $2^n$
saddle point manifolds
\begin{eqnarray}
    \label{sp-solution-fer}
    {\boldsymbol {\mathcal Q}}_n^{\rm (sp)} = -i {\mathfrak s}_z \sqrt{\frac{N}{2}}
    \,{\rm diag}(e^{i \kappa_1\theta},\dots, e^{i \kappa_{n} \theta}),\;\;\;\kappa_\ell=\pm 1,
\end{eqnarray}
are accessible adding up, for $n=2m$, to
\begin{eqnarray}
    \label{Z-F-sp} \fl
    \langle {\mathcal Z}_{2m}^{\rm{(+)}}(z)\rangle
    \simeq\left(  \frac{N}{2} \right)^{mN}
    e^{mN \cos \,2\theta}
    \sum_{q=-m}^{q=+m} {\mathcal V}_{m,q}
    \left(
        \frac{N}{2\pi}
    \right)^{m^2-q^2}
     (2 \sin \, \theta)^{m^2-3q^2}
    \nonumber \\
    \qquad \qquad \qquad \times
    \exp \left[
        iq \left(
            N(2\theta - \sin 2\theta)
            + 2m \left( \theta - \frac{\pi}{2} \right)
        \right)
    \right].
\end{eqnarray}
Here, ${\mathcal V}_{m,q}$ denotes the volume of Grassmanian
\cite{KM-1999b,Z-1999}
\begin{eqnarray} \fl
    \label{Gr-Vol-Barnes}
    {\mathcal V}_{m,q} = {\rm vol}\left[
        \frac{\texttt{U}(2m)}{\texttt{U}(m-q) \times
\texttt{U}(m+q)}
    \right] = (2 \pi)^{m^2-q^2} \frac{\prod_{j=1}^{m+q} \Gamma (j)\, \prod_{j=1}^{m-q} \Gamma (j)}
    {\prod_{j=1}^{2m} \Gamma (j)}.
\end{eqnarray}
The summation index $q$ in Eq.~(\ref{Z-F-sp}) counts $(2m+1)$
families of saddle point manifolds [Eq.~(\ref{sp-solution-fer})],
the $q$-th family being represented by the configuration
\begin{equation}
    \label{f-asym-spm}
    {\boldsymbol {\mathcal Q}}_n^{\rm (sp)}[q] = -i {\mathfrak s}_z \sqrt{\frac{N}{2}}
        \, {\rm diag} (e^{i\theta} \otimes \mathds{1}_{m-q}, \;
    e^{-i\theta} \otimes \mathds{1}_{m+q})
\end{equation}
taken with the obvious combinatorial weight
\begin{eqnarray}
\left( 2m \atop{m+q} \right) =
\frac{\Gamma(2m+1)}{\Gamma(m+1-q)\Gamma(m+1+q)}. \nonumber
\end{eqnarray}
Obtained in the large-$N$ limit, Eq.~(\ref{Z-F-sp}) holds for $m \in
{\mathbb Z}^+$ which do not scale with $N$.

Aimed at deriving the density of eigenlevels through the replica
limit, one should first to analytically continue Eq.~(\ref{Z-F-sp})
away from $m \in {\mathbb Z}^+$. Even though making an analytic
continuation based on an {\it approximate} result is a dangerous
ploy and is, with certainty, a mathematically questionable
procedure, we embark on the proposal due to \cite{KM-1999b} who have
spotted that the volume of Grassmanian [Eq.~(\ref{Gr-Vol-Barnes})]
vanishes for all integers $|q| \ge m+1$. This observation makes it
tempting to extend the summation over $q$ in Eq.~(\ref{Z-F-sp}) to
(minus and plus) infinities to end up with the following trial
function for `analytically continued' fermionic replica partition
function
\begin{eqnarray} \fl
    \label{Z-F-sp2}
\langle {\mathcal Z}_{2m}^{\rm{(+)}}(z)\rangle
    \stackrel{?}\simeq\left(  \frac{N}{2} \right)^{mN}
    e^{mN \cos \,2\theta}
    \sum_{q=-\infty}^{q=+\infty} {\mathcal V}_{m,q}
    \left(
        \frac{N}{2\pi}
    \right)^{m^2-q^2}
     (2 \sin \, \theta)^{m^2-3q^2}
    \nonumber \\
    \qquad \qquad \qquad \times
    \exp \left[
        iq \left(
            N(2\theta - \sin 2\theta)
            + 2m \left( \theta - \frac{\pi}{2} \right)
        \right)
    \right].
\end{eqnarray}
A close inspection of this result reveals that it is flawed: (i)
so-continued replica partition function $ \langle {\mathcal
Z}_{2m}^{\rm{(+)}}(z)\rangle$ diverges~\footnote[6]{The group volume
${\mathcal V}_{m,q}$ grows too fast with $q$ for the series
$\sum_{q=-\infty}^{q=+\infty}(\cdots)$ to converge.}
\cite{KM-1999b,Z-1999} in the vicinity of $m=+0$, the region
crucially important for retrieving the density of eigenlevels
through the replica limit; (ii) due to the $q \mapsto -q$ symmetry
of the summand, $ \langle {\mathcal Z}_{2m}^{\rm{(+)}}(z)\rangle$
must be {\it real} thus leaving no room for a non-zero
density of states. Indeed, a formally derived small-$m$ expansion of
Eq.~(\ref{Z-F-sp2})
\begin{eqnarray}
    \label{small-m-exp} \fl
    \langle {\mathcal Z}_{2m}^{\rm{(+)}}(z)\rangle
    \stackrel{?}\simeq
    1 + m \left[ N
        \left(
            \cos 2\theta + \log \frac{N}{2}
        \right)
        + \frac{1}{4N(\sin \theta)^3}        \, \cos
        \left[
        N(2\theta - \sin 2\theta)
        \right]
    \right] + {\cal O}(m^2)\nonumber\\
    {}
\end{eqnarray}
considered together with Eq.~(\ref{doe}) leads us to conclude that
the replica limit [Eq.~(\ref{fermi-res-av})] in the above
elaboration fails to reproduce even the smooth part of the average
density of eigenlevels yielding $\varrho(\epsilon)= 0$. Notice that
the derivation of Eq.~(\ref{small-m-exp}) boldly ignores the
divergence of the infinite series Eq.~(\ref{Z-F-sp2}).

This unphysical result is at odds with the work \cite{KM-1999b}
where both the Wigner semicircle and the $1/N$ oscillating
correction to it,
\begin{equation}
    \label{z2}
    \varrho(\epsilon)\simeq \frac{2}{\pi} \sqrt{1-\epsilon_{\rm s}^2}
    \left[
        1 - \frac{1}{4N}
        \, \frac{\cos \left(N (2 \theta -\sin 2\theta)
        \right)}{\sin^3 \theta}
    \right],
\end{equation}
were reproduced out of fermionic replicas in almost the same
elaboration. The only difference between the above treatment and
that in \cite{KM-1999b} is that its authors used an {\it alternative
enumeration} of saddle point manifolds Eq.~(\ref{f-asym-spm})
contributing the fermionic replica partition function.

To meet the parameterisation of \cite{KM-1999b}, we introduce a new
summation index, $p = q+m$, in Eq.~(\ref{Z-F-sp}). This amounts to
counting saddle point manifolds starting with the `replica
symmetric' one,
\begin{equation}
\label{spm2}
    \tilde{{\boldsymbol {\mathcal Q}}}_n^{\rm (sp)}[p] = -i {\mathfrak s}_z \sqrt{\frac{N}{2}}
        \, {\rm diag} (e^{i\theta} \otimes \mathds{1}_{2m-p}, \;
    e^{-i\theta} \otimes \mathds{1}_{p})
\end{equation}
as $p$ varies from $0$ to $2m$, so that
\begin{eqnarray}
\label{z2-another}
\fl
    \langle \tilde{{\mathcal Z}}_{2m}^{\rm{(+)}}(z)\rangle
    \simeq
    \left(  \frac{N}{2} \right)^{mN}
    \frac{e^{mN \cos \,2\theta}}{(2 \sin \theta)^{2m^2}}
    \sum_{p=0}^{2 m}{\mathcal V}_{m,\,p-m}
    \left( \frac{N}{2\pi} \right)^{p(2m-p)} \, (2 \sin \, \theta)^{3p(2m-p)}
    \nonumber \\
    \qquad \qquad\qquad\times
    \,
    \exp \left[
        i(p-m) \left( N(2\theta - \sin 2\theta)
            + 2m \left( \theta - \frac{\pi}{2} \right)
        \right)
    \right].
\end{eqnarray}
For $m \in {\mathbb Z^+}$, Eqs.~(\ref{Z-F-sp}) and
(\ref{z2-another}) are trivially identical. However, this is {\it
not} the case for generic real valued $m$ after
Eq.~(\ref{z2-another}) is `analytically continued' by extending the
summation over $p$ to (minus and plus) infinities:
\begin{eqnarray}
\label{z2-another-cont}
\fl
    \langle \tilde{{\mathcal Z}}_{2m}^{\rm{(+)}}(z)\rangle
    \stackrel{?}\simeq
    \left(  \frac{N}{2} \right)^{mN}
    \frac{e^{mN \cos \,2\theta}}{(2 \sin \theta)^{2m^2}}
    \sum_{p=-\infty}^{p=+\infty}{\mathcal V}_{m,\,p-m}
    \left( \frac{N}{2\pi} \right)^{p(2m-p)} \, (2 \sin \, \theta)^{3p(2m-p)}
    \nonumber \\
    \qquad \qquad\qquad\times
    \,
    \exp \left[
        i(p-m) \left( N(2\theta - \sin 2\theta)
            + 2m \left( \theta - \frac{\pi}{2} \right)
        \right)
    \right].
\end{eqnarray}
Now, contrary to the previous result [Eq.~(\ref{small-m-exp})], the
small-$m$ expansion of the fermionic replica partition function
appears to contain a mysterious {\it imaginary} component~\footnote[7]{It was argued in \cite{Z-1999} that the procedure of
`analytic continuation' that led to Eq.~(\ref{z2-another-cont}) and
further to Eq.~(\ref{small-m-exp-2}) favors so-called {\it causal}
saddle points over their conjugate counterparts called {\it
acausal}. Such a selectivity is eventually responsible for the
correct result for the density of eigenlevels as discussed below
Eq.~(\ref{small-m-exp-2}).},
\begin{eqnarray} \fl
    \label{small-m-exp-2}
    \langle \tilde{{\mathcal Z}}_{2m}^{\rm{(+)}}(z)\rangle \stackrel{?}\simeq
    1 + m \left[ N
        \left(
            e^{2i\theta} - 2 i \theta + \log \frac{N}{2}
        \right)
        + \frac{1}{4 N(\sin \theta)^3}\,
        e^{ i N(2\theta - \sin 2\theta)}
    \right]  + {\cal O}(m^2).
\end{eqnarray}
This expansion coincides with the one in \cite{KM-1999b} and does
reproduce, through the replica limit, the correct result
[Eq.~(\ref{z2})] for the average density of eigenvalues in GUE, in
both leading and subleading orders in $1/N$. As the latter term
captures oscillatory behavior of the eigenlevel density, it was
assumed in the literature \cite{KM-1999b,KM-1999c,YL-1999} that the
replica asymmetric saddles may describe a nonperturbative sector of
replica field theories.

Let us stress that neither of the above two treatments of fermionic
replicas [resulting in Eqs.~(\ref{small-m-exp}) and
(\ref{small-m-exp-2}), respectively] can be considered as
mathematically satisfactory because they both rely on a nonexisting
`analytic continuation' of the replica partition function
[Eq.~(\ref{Z-F-sp})] to the vicinity of $m=+0$, as explained below
Eq.~(\ref{Z-F-sp2}).

\newpage
\subsection{Brief summary}
A brief tour d'horizon on the saddle point approach to replica field
theories indicates that the bosonic variation of the replica trick
is restricted to the perturbative sector of the theory accounted for
by the (only reachable) replica symmetric saddle point manifold. In
the fermionic version of the replica trick, both replica symmetric
and replica asymmetric saddle point manifolds contribute the replica
partition function; however, its analytic continuation to a vicinity
of $n=0$ is ill defined, both in terms of convergence and
uniqueness.

These drawbacks of the saddle point approach to replicas are not
specific to the GUE description. Similar difficulties arise in
replica studies of other random matrix ensembles
\cite{DV-2001,NK-2002} and in analysis of physical systems (notably
one-dimensional impenetrable bosons) admitting effective RMT
description \cite{GK-2001,NGK-2003,G-2004}.

Are the problems surfaced in the above calculation indicative of
{\it internal} difficulties of the replica method {\it itself} or
should they be attributed to a particular computational framework? A
little thought shows that the {\it approximate} evaluation of both
bosonic and fermionic replica partition functions is the key point
to blame for the inconsistencies encountered in the above
elaboration of the replica trick. In such a situation, leaning
towards exact calculational schemes in replica field theories is a
natural move.

\section{Integrable theory of replicas}

In this section, we outline an alternative way of treating replica
partition functions. A connection \cite{K-2002,K-2005} between
zero-dimensional replica field theories and the theory of integrable
hierarchies~\footnote[1]{An introductory exposition of integrability arising in the RMT context can be found in Chapter 10 of this Handbook.} is central to our formalism.

\subsection{Density of eigenvalues in the GUE revisited (easy way)}

For illustration purposes~\footnote[8]{Since the framework to be
presented here solely rests on the {\it symmetry} underlying the
matrix model, it can readily be adopted to other spectral statistics
for random matrix ensembles falling into the same $\beta=2$ Dyson's
symmetry class \cite{M-2004}. The reader interested in a general
formulation of the integrable theory of replicas is referred
directly to Section \ref{tau}.}, we choose the very same problem of
calculating the average density of eigenlevels in the GUE specified
by the probability measure Eq.~(\ref{gue-prob}). For the lack of
space, only fermionic replicas will be considered; a bosonic replica
treatment can be found in the tutorial paper \cite{OK-2009}.

\subsubsection{Replica partition function as Toda Lattice}\noindent\newline\newline
Our claim of exact solvability of the replica model
Eq.~(\ref{fer-pf}) and the models of the same ilk rests on {\it two}
observations. To make the {\it first}, we routinely reduce the
average fermionic partition function Eq.~(\ref{fer-pf}) to the
$n$-fold integral
\begin{eqnarray}
\label{e3.1}
    \langle {\mathcal Z}_n^{(+)}(z)\rangle =
    \int_{{\mathbb R}^n} \prod_{\ell=1}^n d\lambda_\ell\, e^{-\lambda_\ell^2}
    \left(\lambda_\ell - iz\right)^{N}\Delta_n^2({\bm \lambda})
\end{eqnarray}
after diagonalising the Hermitean matrix ${\boldsymbol {\mathcal
Q}}_n = {\boldsymbol {\mathcal U}}_n {\boldsymbol \lambda}
{\boldsymbol {\mathcal U}}_n^\dagger$ by unitary rotation
${\boldsymbol {\mathcal U}}_n \in \texttt{U}(n)$; here ${\boldsymbol
\lambda}$ is a diagonal matrix ${\boldsymbol \lambda} ={\rm diag}
(\lambda_1,\dots,\lambda_n)$ composed of eigenvalues of
${\boldsymbol {\mathcal Q}}_n$, and $\Delta_n({\bm \lambda})$ is the
Vandermonde determinant
\begin{eqnarray}
\Delta_n({\bm \lambda}) = \det [\lambda_\ell^{k-1}]=\prod_{\ell
> k}(\lambda_\ell -\lambda_k)
\end{eqnarray}
induced by the Jacobian of the transformation ${\boldsymbol
{\mathcal Q}}_n \mapsto ({\boldsymbol {\mathcal U}}_n, {\boldsymbol
\lambda})$. Further, making a proper shift of the integration
variables and applying the Andr\'eief-de Bruijn integration formula
\cite{A-1883,dB-1955}
\begin{eqnarray}
    \label{dB-formula} \fl \qquad
    \int_{{\mathbb R}^n} \prod_{\ell=1}^n d\mu(\lambda_\ell) \;
    \det [f_k(\lambda_\ell)]
    \det [g_k(\lambda_\ell)]
    =
    n! \; \det \left[
        \int_{\mathbb R} d\mu(\lambda) f_k(\lambda) \, g_\ell (\lambda)
    \right]
\end{eqnarray}
which holds for benign integration measure $d\mu(\lambda)$, one
derives:
\begin{eqnarray}
\label{e3.2}
    \langle {\mathcal Z}_n^{(+)}(z)\rangle =
    \exp\left[ n z^2\right]
    \det{}_{k,\ell} \left[
        \int_{{\mathbb R}} d\lambda \,\lambda^{N+k+\ell}
        \exp\left( -\lambda^2 - 2i z \lambda
        \right)
    \right].
\end{eqnarray}
The latter is equivalent to the remarkable representation
\begin{eqnarray}
    \label{hd-representation}
    \langle {\mathcal Z}_n^{(+)}(z)\rangle
     = \exp\left[ n z^2\right]
    \; {\tau}_n^{(+)}(z)
\end{eqnarray}
involving the {\it Hankel determinant}
\begin{eqnarray}
    \label{hd}
    \tau_{n}^{(+)}(z) = \det
    \left[ \partial_z^{k+\ell} \,
    \tau_1^{(+)} (z)
    \right]_{k,\ell=0, \cdots, n-1}
\end{eqnarray}
with $\tau_1^{(+)}(z) = e^{-z^2} H_N(z)$ being related to the
Hermite polynomial $H_N(z)$ (see also Chapter 4 of this Handbook). In the above equations, no care was
taken of prefactors which tend to unity in the replica limit.

Consequences of the Hankel-determinant-like representation
Eq.~(\ref{hd-representation}) of the fermionic replica partition
function $\langle {\mathcal Z}_n^{(+)}(z)\rangle$ are far reaching.
As had first been shown by Darboux \cite{D-1972} a century ago, any
set of Hankel determinants meeting the `initial condition'
$\tau_0^{(+)}(z)=1$ (which is indeed the case for Eq.~(\ref{hd}) due
to Eq.~(\ref{hd-representation}) and the normalisation $\langle
{\mathcal Z}_0^{(+)}(z)\rangle=1$) satisfies the equation
\begin{eqnarray}
    \label{toda-eq}
    \tau_n^{(+)}(z) \, \frac{\partial^2}{\partial z^2}\tau_n^{(+)}(z)
    - \left(\frac{\partial}{\partial z}\tau_n^{(+)}(z)\right)^2 = \tau_{n-1}^{(+)}(z)
    \tau_{n+1}^{(+)}(z), \;\;\; n \in {\mathbb Z}^+.
\end{eqnarray}
Equations (\ref{hd-representation}) and (\ref{toda-eq}) taken
together with the known initial conditions for $\tau_0^{(+)}(z)$ and
$\tau_1^{(+)}(z)$ establish a differential recursive hierarchy
between {\it nonperturbative} fermionic replica partition functions
$\langle {\mathcal Z}_n^{(+)}(z)\rangle$ with different replica
indices $n$. This is an {\it exact} alternative to the {\it
approximate} solution Eq.~(\ref{Z-F-sp}) presented in the previous
section.

Equation (\ref{toda-eq}), known as the {\it positive Toda Lattice}
equation \cite{T-1967} in the theory of integrable hierarchies~\footnote[7]{See also Chapter 10 of this Handbook.}
\cite{M-1994}, is the first indication of exact solvability
of replica field theories. Importantly, the emergence of the Toda
Lattice hierarchy is eventually due to the $\beta=2$ symmetry of the
fermionic replica field theory encoded into the squared Vandermonde
determinant in Eq.~(\ref{e3.1}).

\subsubsection{From Toda Lattice to Painlev\'e transcendent}\noindent\newline\newline
While important from conceptual point of view, the positive Toda
Lattice equation for the {\it fermionic} replica partition function
$\langle {\mathcal Z}_n^{(+)}(z)\rangle$, if taken alone, is not
much helpful in performing the replica limit.

Fortunately, here the {\it second observation}, borrowed from
\cite{FW-2001}, comes in. Miraculously, the same Toda Lattice
equation governs the behaviour of so-called $\tau$-functions arising
in the Hamiltonian formulation \cite{O-1986} of the six Painlev\'e
equations~\footnote[9]{See also Chapter 9 of this Handbook.} \cite{C-2003,N-2004}, which are yet another
fundamental object in the theory of nonlinear integrable systems.
The Painlev\'e equations contain the hierarchy (or replica) index
$n$ as a {\it parameter}. For this reason, they serve as a proper
starting point \cite{K-2002} for building a consistent analytic
continuation of replica partition functions away from $n$ integers.

The aforementioned Painlev\'e reduction \cite{O-1986,FW-2001} of the
Toda Lattice equation Eq.~(\ref{toda-eq}) materialises in the exact
representation \cite{FW-2001}
\begin{eqnarray}
    \label{Zn-after-GUE-DOS}
    \langle {\mathcal Z}_n^{(+)}(z)\rangle
    = \langle {\mathcal Z}_n^{(+)}(0)\rangle
    \exp \left(
        \int_0^{i\epsilon} dt \, \sigma_{\rm IV}(t)
    \right)
\end{eqnarray}
which holds as soon as $n \in {\mathbb Z}^+$. It involves the fourth
Painlev\'e transcendent $\sigma_{\rm IV}(t)$ satisfying the
Painlev\'e IV equation~\footnote[3]{In the original paper
\cite{K-2002}, Eq.~(\ref{toy-eq-1}) appears to have incorrect signs
in front of $n$ and $N$. I thank
    Nicholas Witte for bringing this fact to my attention.
} in the Jimbo-Miwa-Okamoto form \cite{JM-1981,O-1986}
\begin  {equation}
        \label{toy-eq-1}
        (\sigma_{\rm IV}^{\prime\prime})^2 - 4 (t \sigma_{\rm IV}^\prime
        - \sigma_{\rm IV})^2
        + 4 \sigma_{\rm IV}^\prime (\sigma_{\rm IV}^\prime + 2n)
        (\sigma_{\rm IV}^\prime - 2 N)
        = 0.
\end    {equation}
The boundary condition is $ \sigma_{\rm IV}(t) \sim (nN/t) \left(1 +
{\cal O}(t^{-1}) \right) $ as  $t \rightarrow +\infty$. Note that
Eq.~(\ref{toy-eq-1}), and therefore Eq.~(\ref{Zn-after-GUE-DOS}),
contain the replica index $n$ as a {\it parameter}.

By derivation, Eq.~(\ref{Zn-after-GUE-DOS}) holds for $n$ positive
integers only and, strictly speaking, there is no {\it a priori}
reason to expect it to stay valid away from $n \in {\mathbb Z}^+$.
It can be shown, however, that it {\it is} legitimate to extend
Eq.~(\ref{Zn-after-GUE-DOS}), as it stands, beyond $n \in {\mathbb
Z}^+$ and consider this extension as a {\it proper analytic
continuation} to $n\in {\mathbb R}^+$ we are looking for (the reader
is referred to \cite{K-2002} for a detailed discussion).

As the result, the fermionic replica limit Eq.~(\ref{fermi-res-av})
can now safely be implemented to bring, via Eq.~(\ref{doe}), the
average density of eigenlevels \cite{MG-1960}
\begin{eqnarray}
    \label{m-res}
    \varrho(\epsilon) = \frac{1}{2^{N} \Gamma(N) \sqrt{\pi}}
    \,e^{-\epsilon^2} \left[
        H_N^\prime(\epsilon) H_{N-1}(\epsilon) -
        H_N(\epsilon) H_{N-1}^\prime(\epsilon)
    \right]
\end{eqnarray}
expressed in terms of Hermite polynomials. This is the famous GUE
result firmly established by other methods \cite{M-2004,G-1991}.
Technically, the derivation of Eq.~(\ref{m-res}) is based on the
small-$n$ expansion of the Hamiltonian representation \cite{N-2004}
of the fourth Painlev\'e transcendent. The details of this somewhat
cumbersome calculation can be found in \cite{OK-2009} where the
nonperturbative result Eq.~(\ref{m-res}) is also re-derived within
the bosonic variation of the replica trick.

\subsubsection{Brief summary}\noindent\newline\newline
The above treatment was largely based on a wealth of `ready-for-use'
results (Andr\'eief-de Bruijn formula, Darboux theorem, and a
connection between the Toda Lattice and Painlev\'e transcendents)
which surprisingly well fitted our goal of a nonperturbative
evaluation of the particular replica partition function
Eq.~(\ref{e3.1}). Since existence of such an `easy way' is clearly
the exception rather than the rule, a regular yet flexible formalism
is needed for a nonperturbative description of a general class of
replica partition functions.

\subsection{The $\tau$ function theory of replicas ($\beta=2$)}\label{tau}
In this section, we outline such a regular formalism
\cite{OK-2007,OK-2009} tailor-made for an exact analysis of both
fermionic and bosonic zero-dimensional replica field theories
belonging to the broadly interpreted $\beta=2$ Dyson symmetry class.

\subsubsection{From replica partition function to $\tau$ function}\noindent\newline\newline
Let us concentrate on the fermionic and/or bosonic zero-dimensional
replica field theories whose partition functions admit the
eigenvalue representation
\begin{eqnarray}\label{rpf-def}
    \langle {\mathcal Z}_n^{(\pm)}({\boldsymbol \varsigma})\rangle = \int_{{\cal D}^{n}} \prod_{\ell=1}^{n}
        d\lambda_\ell\, \Gamma({\boldsymbol \varsigma};\lambda_\ell)\, e^{-V_n(\lambda_\ell)}
    \Delta_{n}^2({\boldsymbol \lambda}), \;\;\; n\in {\mathbb Z}^+.
\end{eqnarray}
Here $V_n(\lambda)$ is a `confinement potential' which may depend on
the replica index $\pm n$; $\Gamma({\boldsymbol \varsigma};\lambda)$
is a function accommodating relevant physical parameters
${\boldsymbol \varsigma}=(\varsigma_1,\varsigma_2,\dots)$ of the
theory (e.g., energies in the multi-point spectral correlation
functions). In order to treat the fermionic and bosonic replicas on
the same footing, the integration domain ${\mathcal D}$ was chosen
to be~\footnote[1]{Notice that ${\mathcal D}=[-1,+1]$ for (compact)
fermionic replicas, and ${\mathcal D}=[0,+\infty)$ for (noncompact)
bosonic replicas. A more general setting Eq.~(\ref{domain}) does not
complicate the theory.}
\begin{eqnarray} \label{domain}
{\mathcal D}=\bigcup_{j=1}^r [c_{2j-1},c_{2j}].
\end{eqnarray}

To determine the replica partition function $\langle {\mathcal
Z}_n^{(\pm)}({\boldsymbol \varsigma})\rangle$ nonperturbatively, we
adopt the `deform-and-study' approach, a standard string theory
method of revealing hidden structures. Its main idea consists of
`embedding' $\langle {\mathcal Z}_n^{(\pm)}({\boldsymbol
\varsigma})\rangle$ into a more general theory of $\tau$ functions
\begin{eqnarray}
\label{tau-f}
    \tau_{n}^{(s)}({\boldsymbol \varsigma}; {\boldsymbol t}) &=& \frac{1}{n!}
    \int_{{\mathcal D}^{n}} \prod_{\ell=1}^{n}
        d\lambda_\ell\,  \Gamma({\boldsymbol \varsigma};\lambda_\ell) \, e^{-V_{n- s}(\lambda_\ell)} \,
        e^{v({\boldsymbol t};\lambda_k)} \,
     \Delta_{n}^2({\boldsymbol \lambda})
\end{eqnarray}
which posses the infinite-dimensional parameter space ${\bm t}=(t_1,
t_2,\cdots)$ arising as the result of the ${\boldsymbol
t}$-deformation
\begin{eqnarray}
v({\boldsymbol t};\lambda) = \sum_{j=1}^\infty t_j \lambda^j
\end{eqnarray}
of the confinement potential. The auxiliary parameter $s$ is assumed
to be an integer, $s \in {\mathbb Z}$. Studying the evolution of
$\tau$ functions in the extended $(n,s,{\bm t}, {\bm \varsigma})$
space allows us to identify the highly nontrivial, nonlinear
differential hierarchical relations between them. Projection of
these relations, taken at $s=0$, onto the hyperplane ${\bm t}={\bm
0}$,
\begin{eqnarray}
\label{proj}
    \langle {\mathcal Z}_n^{(\pm)}({\boldsymbol \varsigma})\rangle  = n!\,\tau_n^{(s)}({\bm \varsigma}; {\bm t})
    \Big|_{s=0,\,{\bm t}={\bm 0}},
\end{eqnarray}
will generate, among others, a closed nonlinear differential
equation for the replica partition function $\langle {\mathcal
Z}_n^{(\pm)}({\boldsymbol \varsigma})\rangle $. Since this {\it
nonperturbative} equation appears to contain the replica (or
hierarchy) index $n$ as a parameter, it is expected \cite{K-2002} to
serve as a proper starting point for building a consistent analytic
continuation of $\langle {\mathcal Z}_n^{(\pm)}({\boldsymbol
\varsigma})\rangle$ away from $n$ integers.

Having formulated the crux of the method, let us turn to its
exposition. The two key ingredients of the exact theory of $\tau$
functions are (i) the bilinear identity \cite{DKJM-1983,ASvM-1995}
and (ii) the (linear) Virasoro constraints \cite{MM-1990}.

\subsubsection{Bilinear identity and integrable hierarchies}\noindent\newline\newline
The bilinear identity encodes an infinite set of hierarchically
structured nonlinear differential equations in the variables
$\{t_j\}$. For the model introduced in Eq.~(\ref{tau-f}), the
bilinear identity reads \cite{ASvM-1995,TSY-1996,OK-2007,OK-2009}:
\begin{eqnarray} \label{bi-id} \fl
    \qquad \oint_{{\cal C}_\infty} dz \,e^{a\, v(\bm{t-t^\prime};z)}\Bigg(
    \tau_{n}^{(s)}(\bm{t}-[\bm{z}^{-1}])\,
    \frac{\tau_{m+1}^{(m+1+s-n)}(\bm{t^\prime}+[\bm{z}^{-1}])}{z^{m+1-n}}\, e^{v(\bm{t-t^\prime};z)}
    \nonumber\\
    \qquad \qquad \qquad
    -\tau_m^{(m+s-n)}(\bm{t^\prime}-[\bm{z}^{-1}]) \frac{\tau_{n+1}^{(s+1)}
    (\bm{t}+[\bm{z}^{-1}])}{z^{n+1-m}}\,
    \Bigg)=0.
\end{eqnarray}
Here, $a\in {\mathbb R}$ is a free parameter; the integration
contour ${\cal C}_\infty$ encompasses the point $z=\infty$; the
notation ${\bm t} \pm [{\bm z}^{-1}]$ stands for the infinite set of
parameters $\{t_j\pm z^{-j}/j\}$; for brevity, the physical
parameters ${\bm \varsigma}$ were dropped from the arguments of
$\tau$ functions.

Being expanded in terms of $\bm{ t^\prime}-{\bm t}$ and $a$,
Eq.~(\ref{bi-id}) generates various integrable hierarchies. One of
them, the Kadomtsev-Petviashvili (KP) hierarchy in the Hirota
form~\footnote[2]{In Eq.~(\ref{kph}), the $j$-th component of the
infinite-dimensional
    vector $[{\bm D}]$ equals $j^{-1}D_j$; the functions $s_k({\bm t})$ are the Schur
    polynomials \cite{MD-1998} defined by the expansion
    \begin{eqnarray}
        \exp\left( \sum_{j=1}^\infty t_j x^j \right) = \sum_{k=0}^\infty
    x^k s_k({\bm t}).
    \nonumber
    \end{eqnarray}
The operator symbol $D_j \,f({\bm t})\circ g({\bm t})$ stands for
the
    Hirota derivative
    $\partial_{x_j} \,f({\bm t}+{\bm x})\,g({\bm t}-{\bm x})\,\big|_{{\bm x}=\bm{0}}$.}
\begin{equation}
   \label{kph}
    \frac{1}{2}\,D_1 D_k\, \tau_n^{(s)}(\bm{t})
    \circ \tau_n^{(s)}(\bm{t}) =  s_{k+1}([\bm{D}]) \, \tau_n^{(s)}(\bm{t})
    \circ \tau_n^{(s)}(\bm{t})
\end{equation}
($k\ge 3$) is of primary importance for the exact theory of
replicas. The first nontrivial member of the KP hierarchy reads
\begin{eqnarray} \fl
    \label{fkp}
    \left(
    \frac{\partial^4}{\partial t_1^4} + 3\,\frac{\partial^2}{\partial t_2^2} -
        4\, \frac{\partial^2}{\partial t_1  \partial t_3}
    \right)\, \log \tau_n^{(s)}(\bm{\varsigma};{\bm t})
    + \,6\, \left(
        \frac{\partial^2}{\partial t_1^2}\, \log \tau_n^{(s)}(\bm{\varsigma};{\bm t})
    \right)^2 = 0.
\end{eqnarray}
In what follows, it will be shown that its projection onto $s=0$ and
${\bm t}={\bm 0}$ [Eq.~(\ref{proj})] gives rise to a nonlinear
differential equation for the replica partition function $\langle
{\mathcal Z}_n^{(\pm)}({\boldsymbol \varsigma})\rangle$.

\subsubsection{Virasoro constraints}\noindent\newline\newline
Since we are interested in deriving a differential equation for
$\langle {\mathcal Z}_n^{(\pm)}({\boldsymbol \varsigma})\rangle$ in
terms of the derivatives over {\it physical parameters}
$\{\varsigma_j\}$, we have to seek an additional block of the theory
that would make a link between the $\{t_j\}$ derivatives in
Eq.~(\ref{fkp}) taken at ${\bm t}={\bm 0}$ and the derivatives over
physical parameters $\{\varsigma_j\}$. The study \cite{ASvM-1995} by
Adler, Shiota, and van Moerbeke suggests that the missing block is
the {\it Virasoro constraints}~\footnote[7]{See also Chapter 10 of this Handbook.} which reflect the invariance of the
$\tau$ function [Eq.~(\ref{tau-f})] under a change of the
integration variables.

In the present context, it is useful to demand the invariance under
an infinite set of transformations
\begin{eqnarray}\label{vt}
    \lambda_j \rightarrow \mu_j + \epsilon \mu_j^{q+1} f(\mu_j)
    \prod_{k=1}^{\mathfrak{m}} (\mu_j - c_k^\prime),\;\;\;
    q \ge -1,
\end{eqnarray}
labeled by integers $q$. Here, $\epsilon>0$, the vector
$\bm{c^\prime}$ is ${\bm
c}^\prime=\{c_1,\cdots,c_{2r}\}\setminus\{\pm \infty, {\bm
\aleph}\}$ with $\bm{\aleph}$ denoting a set of zeros of
$f(\lambda)$, and $\mathfrak{m} = {\rm dim\,}(\bm{c^\prime})$. The
function $f(\lambda)$ is, in turn, related to the confinement
potential $V_{n-s}(\lambda)$ through the parameterisation
\begin{eqnarray}
\label{vns} \frac{dV_{n-s}}{d\lambda} =
\frac{g(\lambda)}{f(\lambda)},\;\;\;g(\lambda)=\sum_{k=0}^\infty b_k
\lambda^k,\;\;\; f(\lambda)=\sum_{k=0}^\infty a_k \lambda^k
\end{eqnarray}
in which both $g(\lambda)$ and $f(\lambda)$ depend on $n-s$ as do
the coefficients $b_k$ and $a_k$ in the above expansions. The
transformation Eq.~(\ref{vt}) induces the Virasoro constraints
\cite{OK-2007}
\begin{equation}
\label{2-Vir}
    \left[ \hat{{\cal L}}_{q}^V({\bm t}) + \hat{{\cal L}}_q^{\Gamma}({\bm \varsigma};{\bm t})
     \right] \tau_n^{(s)}({\bm \varsigma};{\bm t})=0,
\end{equation}
where the differential operator
\begin{eqnarray} \fl
\label{vLv}
     \hat{{\cal L}}_{q}^V({\bm t}) = \sum_{\ell = 0}^\infty
    \sum_{k=0}^{\mathfrak{m}}  s_{\mathfrak{m}-k}(-{\bm p}_\mathfrak{m} (\bm{c^\prime}))
       \left(
        a_\ell \hat{\cal L}_{q+k+\ell}({\bm t}) - b_\ell \frac{\partial}{\partial t_{q+k+\ell+1}}
    \right),
\end{eqnarray}
acting in the ${\bm t}$-space, is expressed in terms of the Virasoro
operators~\footnote[3]{Equation (\ref{vo}) assumes that
$\partial/\partial {t_0}$ is identified with the multiplicity of the
matrix integral in Eq.~(\ref{tau-f}), $\partial/\partial{t_0} \equiv
n$.}
\begin{eqnarray}
    \label{vo}
    \hat{{\cal L}}_q({\bm t}) = \sum_{j=1}^\infty jt_j \,\frac{\partial}{\partial t_{q+j}}
    +
    \sum_{j=0}^q \frac{\partial^2}{\partial {t_j}\partial {t_{q-j}}},
\end{eqnarray}
obeying the Virasoro algebra
\begin{eqnarray}
\label{va}
[\hat{{\cal L}}_p,\hat{{\cal L}}_q] = (p-q)\hat{{\cal L}}_{p+q}, \;\;\;
p,q\ge -1.
\end{eqnarray}
The notation $s_k(-{\bm p}_{\mathfrak m} (\bm{c^\prime}))$ stands
for the Schur polynomial and ${\bm p}_{\mathfrak m}(\bm{c^\prime})$
is an infinite dimensional vector
\begin{eqnarray} \label{b9090} \fl
    {\bm p}_{\mathfrak m}(\bm{c^\prime})=\left(
    {\rm tr}_{\mathfrak m}(\bm{c^\prime}), \frac{1}{2} {\rm tr}_{\mathfrak m}(\bm{c^\prime})^2,\cdots,
    \frac{1}{k} {\rm tr}_{\mathfrak m}(\bm{c^\prime})^k,\cdots
    \right),
\end{eqnarray}
where ${\rm tr}_{\mathfrak m}(\bm{c^\prime})^k
=\sum_{j=1}^{\mathfrak m} (c_j^\prime)^k$.

While very similar in spirit, the calculation of $\hat{{\cal
L}}_{q}^\Gamma({\bm t})$, the second ingredient in
Eq.~(\ref{2-Vir}), is more of an art since the function $\Gamma({\bm
\varsigma};\lambda)$ in Eq.~(\ref{tau-f}) may significantly vary
from one replica model to the other.

\subsubsection{From $\tau$ function to replica partition function}\noindent\newline\newline
Remarkably, for ${\boldsymbol{t}={\boldsymbol 0}}$, the two
equations [Eqs. (\ref{fkp}) and (\ref{2-Vir})] can be solved jointly
to bring a closed nonlinear differential equation for $\langle
{\mathcal Z}_n^{(\pm)}({\boldsymbol \varsigma})\rangle$. It is this
equation which, being supplemented by appropriate boundary
conditions, provides a truly nonperturbative description of the
replica partition functions and facilitates performing the replica
limit.

\subsection{Exact (bosonic) replicas at work: Density of eigenlevels in the chiral GUE}

\subsubsection{Definitions}\noindent\newline\newline
To see the above formalism at work, let us consider the chiral
Gaussian Unitary Ensemble (chGUE) of $N\times N$ random matrices
\begin{eqnarray}
\label{dm}
    {\boldsymbol {\mathcal H}}_{\mathcal D} = \left(
                 \begin{array}{cc}
                   0 & {\boldsymbol {\mathcal W}} \\
                   {\boldsymbol {\mathcal W}}^\dagger & 0 \\
                 \end{array}
               \right)
\end{eqnarray}
known to describe the low-energy sector of $\texttt{SU}(N_c\ge 3)$
QCD in the fundamental representation~\footnote[5]{See also Chapter 32 of this Handbook.} \cite{VZ-1993}.
Composed of rectangular $n_L\times n_R$ random matrices
${\boldsymbol {\mathcal W}}$ with the Gaussian distributed complex
valued entries
\begin{eqnarray}
\label{dm-den}
    P_{n_L,n_R}({\boldsymbol {\mathcal W}}) =\left(
     \frac{2\pi}{N\Sigma^2}
    \right)^{n_L n_R}
     \exp\left[-\frac{N\Sigma^2}{2}{\rm tr\,} {\boldsymbol {\mathcal W}}^\dagger {\boldsymbol {\mathcal W}}\right],
\end{eqnarray}
where $N=n_L + n_R$, the matrix ${\boldsymbol {\mathcal
H}}_{\mathcal D}$ has exactly $\nu =|n_R-n_L|$ zero eigenvalues
identified with the topological charge $\nu$; the remaining
eigenvalues occur in pairs $\{\pm \lambda_j\}$; the parameter
$\Sigma$ denotes the chiral condensate.

\subsubsection{Nonperturbative calculation of bosonic replica partition function}\noindent\newline\newline
To determine the (microscopic) spectral density from the bosonic
replicas, we define the replica partition function
\begin{eqnarray}
\langle {\mathcal Z}_n^{(-)}(\varsigma)\rangle = \left< {\rm
det}^{-n}(\varsigma + i{\boldsymbol {\mathcal H}}_{\mathcal
D})\right>_{\boldsymbol {\mathcal W}}
\end{eqnarray}
(the angular brackets denote averaging with respect to the
probability density Eq.~(\ref{dm-den})) and map it onto a bosonic
field theory. In the half-plane $\mathfrak{Re}\,\varsigma
>0$, the partition function $\langle {\mathcal
Z}_n^{(-)}(\varsigma)\rangle$ reduces to \cite{DV-2001,F-2002}
\begin{equation}
\label{brft}
    {\mathcal Z}_n^{(-)}(\omega) = \int_{{\mathcal S}_n}
    d\boldsymbol{\mathcal{Q}}_n\, \det{}^{\nu-n} \boldsymbol{\mathcal{Q}}_n \, \exp\left[
    -\frac{\omega}{2} {\rm Tr} (\boldsymbol{\mathcal{Q}}_n + \boldsymbol{\mathcal{Q}}_n^{-1})
    \right],
\end{equation}
where the integration domain ${\mathcal S}_n$ spans all $n\times n$
positive definite Hermitean matrices $\boldsymbol{\mathcal{Q}}_n$.
Equation (\ref{brft}) was derived in the thermodynamic limit
$N\rightarrow \infty$ with the spectral parameter $\omega=\varsigma
N\Sigma$ being kept fixed ($\mathfrak{Re}\,\omega>0$).

Spotting the invariance of the integrand in Eq.~(\ref{brft}) under
the unitary rotation of the matrix $\boldsymbol{{\mathcal Q}}_n$,
one readily realises that ${\mathcal Z}_n^{(-)}(\omega)$ belongs to
the class of $\tau$ functions specified by Eq.~(\ref{tau-f}) where
${\cal D}$ is set to ${\mathbb R}^+$, the potential $V_{n-s}$ is
\begin{eqnarray} \label{pot}
V_{n-s}(\lambda) = (n-s-\nu)\, \log \lambda,
\end{eqnarray}
and $\Gamma({\boldsymbol\varsigma};\lambda)$ is replaced with
\begin{eqnarray}\label{gamma}
\Gamma(\omega;\lambda) = \exp\left[-\frac{\omega}{2}
    \left(
            \lambda  + \frac{1}{\lambda}\right)\right].
\end{eqnarray}
This observation implies that the associated $\tau$ function
$\tau_n^{(s)}(\omega; {\bm t})$ satisfies both the first KP equation
(\ref{fkp}) and the Virasoro constraints Eq.~(\ref{2-Vir}) with
\cite{OK-2007}
\begin{eqnarray}
\label{L-qv-bos-1}
   \hat{{\cal L}}_{q}^V({\bm t}) &=& \hat{{\cal L}}_{q+1}({\bm t}) + (\nu-n+s) \,\frac{\partial}{\partial {t_{q+1}}},\\
   \label{L-qv-bos-2}
   \hat{{\cal L}}_{q}^\Gamma(\omega;{\bm t}) &=&
    - \frac{\omega}{2} \frac{\partial}{\partial {t_{q+2}}}
        -\delta_{q,\,-1}
  \left(
        \omega \frac{\partial}{\partial \omega} + \frac{\omega}{2}\,\frac{\partial}{\partial {t_1}}
    \right) +
    \left[1-\delta_{q,\,-1}\right]\, \frac{\omega}{2}\,\frac{\partial}{\partial {t_q}}.
\end{eqnarray}
Projecting Eq.~(\ref{fkp}) taken at $s=0$ onto ${\bm t}={\bm 0}$,
and expressing the partial derivatives therein via the derivatives
over $\omega$ with the help of Eqs.~(\ref{2-Vir}),
(\ref{L-qv-bos-1}) and (\ref{L-qv-bos-2}), we conclude that
\begin{eqnarray}
    h_n(\omega) = \frac{\partial}{\partial \omega} \log {\mathcal Z}_n^{(-)}(\omega)
\end{eqnarray}
obeys the differential equation \cite{OK-2007}
\begin{eqnarray}
\label{fin-eq} \fl
  h_n^{\prime\prime\prime} + \frac{2}{\omega} h_n^{\prime\prime}
  - \left( 4 + \frac{1+4(n^2+\nu^2)}{\omega^2}\right) h_n^{\prime}
  + 6(h_n^{\prime})^2 \nonumber\\
  \qquad \qquad + \frac{1 - 4 (n^2+\nu^2)}{\omega^3} h_n
  -\frac{2}{\omega^2} (h_n)^2 + \frac{4}{\omega} h_n h_n^{\prime}
  +\frac{4n^2}{\omega^2} = 0
\end{eqnarray}
that can be reduced to the Painlev\'e III.

Considered together with the boundary conditions
$h_n(\omega\rightarrow 0) \simeq -n\nu/\omega$ and
$h_n(\omega\rightarrow \infty) \simeq -n - n^2/(2\omega)$, following
from Eq.~(\ref{brft}), the nonlinear differential equation
Eq.~(\ref{fin-eq}) provides a nonperturbative characterisation of the {\it bosonic} replica partition function~\footnote[7]{Interestingly, the {\it fermionic} replica partition function \cite{LS-1992,SV-1995,DV-2001}
$$
   {\mathcal Z}_n^{(+)}(\omega) = \int_{{\boldsymbol {\mathcal U}}_n {\boldsymbol {\mathcal U}}_n^\dagger =\mathds{1}_n}
    {\mathcal D}\boldsymbol{\mathcal{U}}_n\, \det{}^{\nu} \boldsymbol{\mathcal{U}}_n \, \exp\left[
    \frac{\omega}{2} {\rm Tr} (\boldsymbol{\mathcal{U}}_n + \boldsymbol{\mathcal{U}}_n^{\dagger})
    \right],
$$
admits a similar representation in terms of the Painlev\'e III. Specifically, the function\linebreak $f_n(\omega) = (\partial/\partial \omega) \, \log {\mathcal Z}_n^{(+)}(\omega)$ can be shown \cite{K-2002} to satisfy the very same Eq.~(\ref{fin-eq}) supplemented by the boundary conditions $f_n(\omega\rightarrow \omega_0) = h_{-n}(\omega\rightarrow \omega_0)$, where $\omega_0=\{0,\, \infty\}$.}
${\mathcal Z}_n^{(-)}(\omega)$ for all $n\in {\mathbb Z}^+$.

\subsubsection{Implementing the replica limit}\noindent\newline\newline
To pave the way for the replica calculation of the resolvent
$g(\omega)$ determined by the replica limit
\begin{eqnarray}
g(\omega)=-\lim_{n \rightarrow 0} \frac{1}{n}\frac{\partial}{\partial \omega} \log {\mathcal Z}_n^{(-)}(\omega) =
-\lim_{n \rightarrow 0} \frac{1}{n} h_n(\omega),
\end{eqnarray}
one has to analytically continue $h_n(\omega)$ away from $n$
integers. Previous studies \cite{K-2002,K-2005} suggest that the
sought analytic continuation is given by the very same
Eq.~(\ref{fin-eq}) where the replica parameter $n$ is let to explore
the entire real axis. This leap makes the rest of the calculation
straightforward. Representing $h_n(\omega)$ in the vicinity of $n=0$
as\linebreak $h_n(\omega) = \sum_{p=1}^\infty n^p a_p(\omega)$, we
conclude that $g(\omega) = -a_1(\omega)$ satisfies the equation
\begin{equation}
    \omega^3 g^{\prime\prime\prime} + 2 \omega^2 g^{\prime\prime}
  - \left( 1 + 4 \nu^2 + 4 \omega^2\right)\omega g^{\prime}
  + (1 - 4 \nu^2) \,g = 0.
\end{equation}
Its solution, subject to the boundary conditions consistent with
those specified below Eq.~(\ref{fin-eq}),
\begin{eqnarray}
    g(\omega) = \frac{\nu}{\omega} + + \omega\, \Big[
        I_\nu(\omega) \, K_\nu(\omega) + I_{\nu+1}(\omega) \, K_{\nu-1}(\omega)
    \Big],
\end{eqnarray}
brings the microscopic spectral density $\varrho(\omega) = \pi^{-1}
\mathfrak{Re}\, g(i\omega+0)$ in the form
\begin{eqnarray}
\label{den-fin}
    \varrho(\omega) = \nu \delta(\omega) + \frac{\omega}{2} \Big[
        J_\nu^2(\omega) - J_{\nu-1}(\omega) J_{\nu+1}(\omega)
    \Big].
\end{eqnarray}
In the above formula (which is one of the celebrated RMT results
originally obtained in \cite{VZ-1993} within the orthogonal polynomial technique), the
function $J_\nu$ denotes the Bessel function of the first kind,
whilst $I_\nu$ and $K_\nu$ are the modified Bessel function of the
first and second kind, respectively. Let us stress that the
approximate, saddle point approach to bosonic replicas
\cite{DV-2001} fails to produce Eq.~(\ref{den-fin}).

\section{Concluding remarks}
Concluding this brief excursion into integrable theory of replica
field theories, we wish to mention another important development due
to Splittorff and Verbaarschot \cite{SV-2003,SV-2004}, not reviewed
here for a lack of space. These authors showed that nonperturbative
results for various RMT correlation functions at $\beta=2$ can be
derived by taking the replica limit of the {\it graded} Toda Lattice
equation whose positive ($n\in {\mathbb Z}^+$) and negative ($n\in
{\mathbb Z}^-$) branches describe fermionic and bosonic replica
partition functions, respectively. Being a supersymmetric in nature,
this approach greatly simplifies calculations of spectral
correlation functions through a remarkable fermionic-bosonic
factorisation. For further details, the reader is referred to the
original papers \cite{SV-2003,SV-2004}, the lecture notes
\cite{V-2005}, Chapter 32 of this Handbook, and the tutorial paper \cite{OK-2009}. 

Certainly, more effort is needed to accomplish integrable theory of
zero-dimensional replica field theories. In particular, its
extension to the $\beta=1$ and $\beta=4$ Dyson symmetry classes is
very much called for.
\newline\newline\newline\noindent {\bf Acknowledgements}. I am indebted to Vladimir Al.
Osipov for collaboration on the `replica project'. This work was
supported by the Israel Science Foundation through Grants No. 286/04
and No. 414/08.

\newpage
\section*{References}
\fancyhead{} \fancyhead[RE,LO]{References}
\fancyhead[LE,RO]{\thepage}
\addcontentsline{toc}{section}{\protect\enlargethispage*{100pt}References}

\end{document}